\documentclass{optica-article}

\journal{opticajournal} 

\articletype{Research Article}

\usepackage{lineno}
\usepackage{float}

\begin{document}

\title{Simultaneous two-dimensional velocity and distance measurements based on laser triangulation}

\author{Hao Zhang\authormark{*} and Shiji Wang\authormark{}}

\address{\authormark{}Department of Measurement and Control Technology and Instrumentation, Dalian Maritime University, 1 Linghai Road, Dalian, 116026, China}

\email{\authormark{*}hao.zhang@dlmu.edu.cn} 


\begin{abstract*} 
Laser triangulation sensors are widely used in industry for surface inspection due to simple setup, micron precision and low cost. Conventional laser triangulation methods only enable axial distance measurement limiting further applications, and their lateral resolution is limited by surface microstructure. For overcoming these issues, based on the geometric optics we propose novel theoretical models and methods to achieve lateral velocity measurement. Moreover, a novel axial distance measurement method using edge detection is presented, which can increase the lateral resolution by the order of one magnitude. The performance of the proposed methods are validated through simultaneous orthogonal velocity and distance measurements on a moving established metal specimen, showing the relative error and relative uncertainty can reach $10^{-4}$. The versatility of this multi degree of freedom measurement method paves the way for its broad application across all laser triangulation systems. Therefore, this simultaneous two-dimensional velocity and distance sensing approach can propel advancements in dynamic behavior discipline, including but not limited to motion mechanology and fluid mechanics.

\end{abstract*}

\section{Introduction}\label{c1}

Velocity, distance, shape and vibration measurements of moving objects for instance in cutting lathes, production lines and fluids are significant tasks in the field of process and quality monitoring, which are greatly helpful to enhance the performance of systems and products \cite{kuschmierz2016optical,zhang2020non,fischer2009measurements,zhang2022laser}. As the state of the art, coordinate measuring machines (CMMs) enable submicron measurement for surface quality inspection of precision parts \cite{bills2012volumetric,hocken2016coordinate}. However, due to the tactile nature of CMMs the measurements are commonly conducted off-line and ex-situ, which are time-consuming and easily lead to additional clamping deviations. Meanwhile, it has a potential risk of damaging the measured surface and cannot be applied to fluids. Therefore, non-contact, in-situ measurement methods are required.  

Optical measurement techniques are contactless and efficient, which allow measurement rates of kilohertz level and micron or even submicron precision. These are significant for the measurements of sensitive and fast moving surfaces. Because of the above advantages a lot of optical sensors are applied for various measurement purposes. Confocal sensors are commonly employed for displacement measurements within a working range of millimeters, offering submicron precision \cite{jordan1998highly,tao2011adaptive}. Time-of-Flight (TOF) \cite{horaud2016overview,amann2001laser} employs a rangefinder to emit light pulses and calculates the round-trip time or phase delay of the light waves between the object and the measuring instrument to determine distance. But they can only achieve millimeter uncertainty currently. Low coherence interferometry (LCI) \cite{dufour2005low}, also known as white light interferometry (WLI) \cite{schnell1996dispersive,zhang2019fiber} or optical coherence tomography (OCT) \cite{huang1991optical}, is commonly used for distance and displacement measurements in the sub-millimeter range, achieving a measurement uncertainty of approximately one micron. As an enhanced interferometric technique, synthetic wavelength interferometry can measure the shape of objects with large step heights \cite{lu2002measuring}. Moreover, through the utilization of digital cameras, both holography and shearography have rapidly advanced for submicron-precision distance and 3-D measurements \cite{yang1995precision,yamaguchi1997phase}. Structured light techniques can achieve full-field 3-D measurements with micron-level uncertainties by projecting fringes \cite{tang1990fast} or other patterns \cite{morita1988reconstruction,maruyama1993range} onto the object surface. Utilizing stereo vision, digital image correlation (DIC) \cite{sutton1983determination,wang2010whole} enables micron precision in both in-plane and out-of-plane displacement measurements by correlating the speckle patterns from either spray or laser on the measured surface. At present, several of these techniques have been used for in-situ measurements \cite{frade2012situ,zou2017non} and even for in-line measurements \cite{asundi2006time,strohmeier2019optical}. Nevertheless, all these techniques can provide only one measurand, namely distance, which is not sufficient for many application requirements for instance for determining the absolute shape of a rotating object.

As one of the most commonly utilized optical measurement techniques, laser triangulation enables distance measurement through the geometric principles of similar triangles providing micron-level uncertainty and a simple setup \cite{ji1989design,dorsch1994laser,nan2022automatic}. However, in the past few decades it only enables axial ranging as well, which limits its further application. Besides, the widely used centroid position based ranging method of laser triangulation systems suffers from the interference of surface microstructure, restricting the measurement resolution. 

In order to overcome above issues and achieve a breakthrough in the development of laser triangulation, a laser triangulation based simultaneous lateral velocity and axial distance measurement method is proposed and investigated in this paper. Firstly, the theoretical models and methods for the velocity and distance measurements are presented based on the geometric optics, speckle effect and image processing in Section \ref{c2}. Furthermore, developing a laser triangulation system Section \ref{c3} conducts experimental investigations and demonstrates the effectiveness of the proposed methods. Finally, in Section \ref{c4} the results are discussed and concluded showing potential impacts of significance.

\section{Principle and method}\label{c2}

\subsection{System structure}\label{subsec1}
\begin{figure}[H]
\centering
\includegraphics[width=0.8\textwidth]{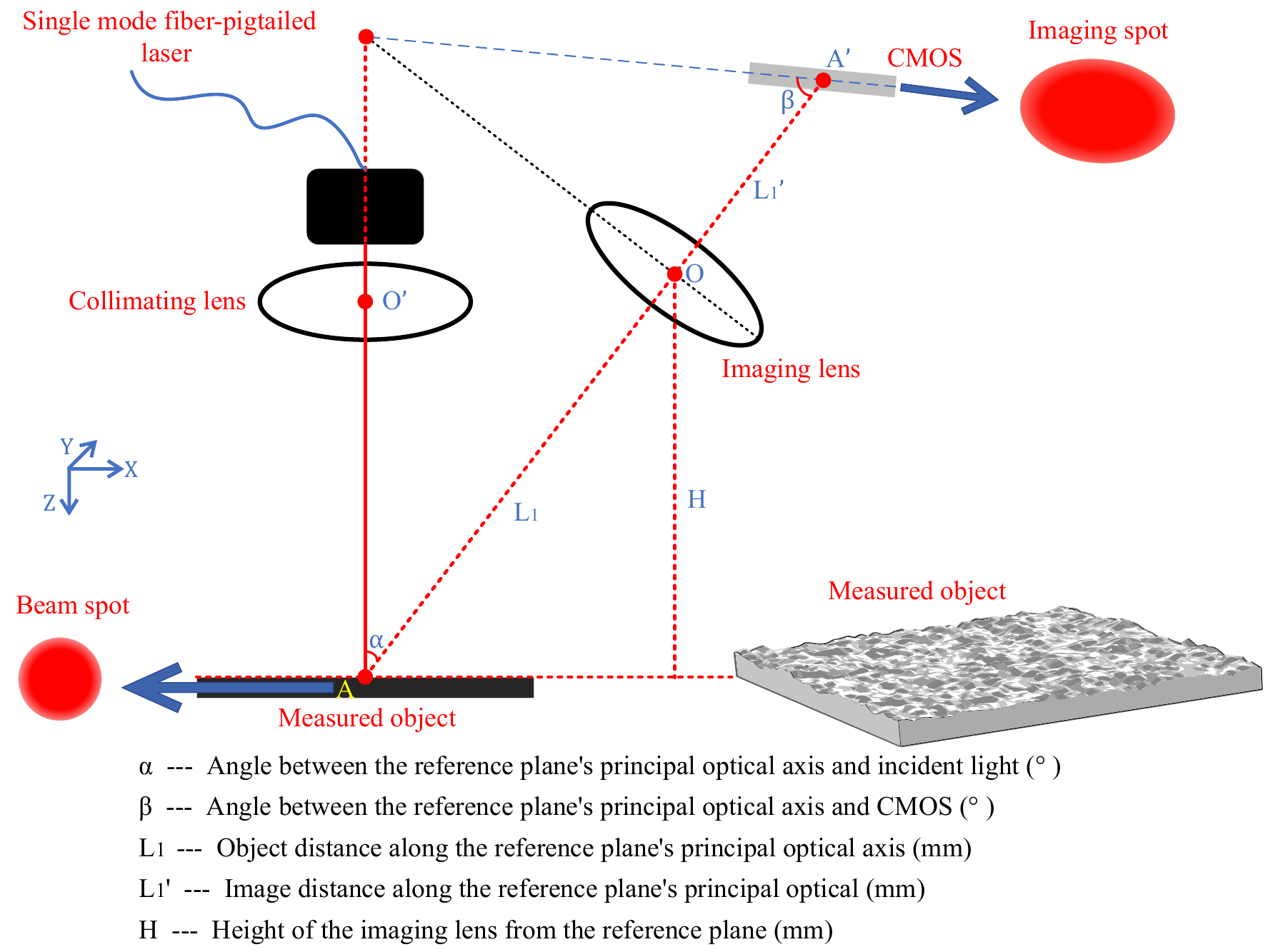}
\caption{Optical structure of the laser triangulation system.}\label{f1}
\end{figure}

The optical structure of a laser triangulation system is shown in Fig. \ref{f1}. 
The system uses a semiconductor laser to emit a Gaussian beam. The beam is passing a collimating lens and then projected vertically onto a measured rough surface, forming a beam spot and diffuse reflection. The scattered light leads to the speckle effect \cite{goodman2007speckle,zhang2016camera,zhang2018miniaturized}, which is received by an imaging lens and finally imaged on a complementary metal oxide semiconductor (CMOS) device producing speckle pattern. When the measured surface moves axially ($z$ direction) or laterally ($x$ direction), the speckle pattern bodily moves or partially changes, respectively.

Focusing scattered light through the imaging lens is one of the core processes for laser triangulation sensing. In order to keep the imaging spot clear on the CMOS during the motion of measured surface, the system structure of the laser triangulation sensor should obey Scheimpflug rule \cite{miks2013analysis}. The incident direction of the beam, the main plane of the imaging lens, and the extension line of the CMOS intersect at one line, which can be expressed by 
\begin{equation}\label{e1}
L_1\cdot\tan{\alpha}=L_1^\prime\cdot\tan{\beta}
\end{equation}

\subsection{Lateral velocity measurement}\label{subsec2}
\begin{figure}[H]
\centering
\includegraphics[width=1\textwidth]{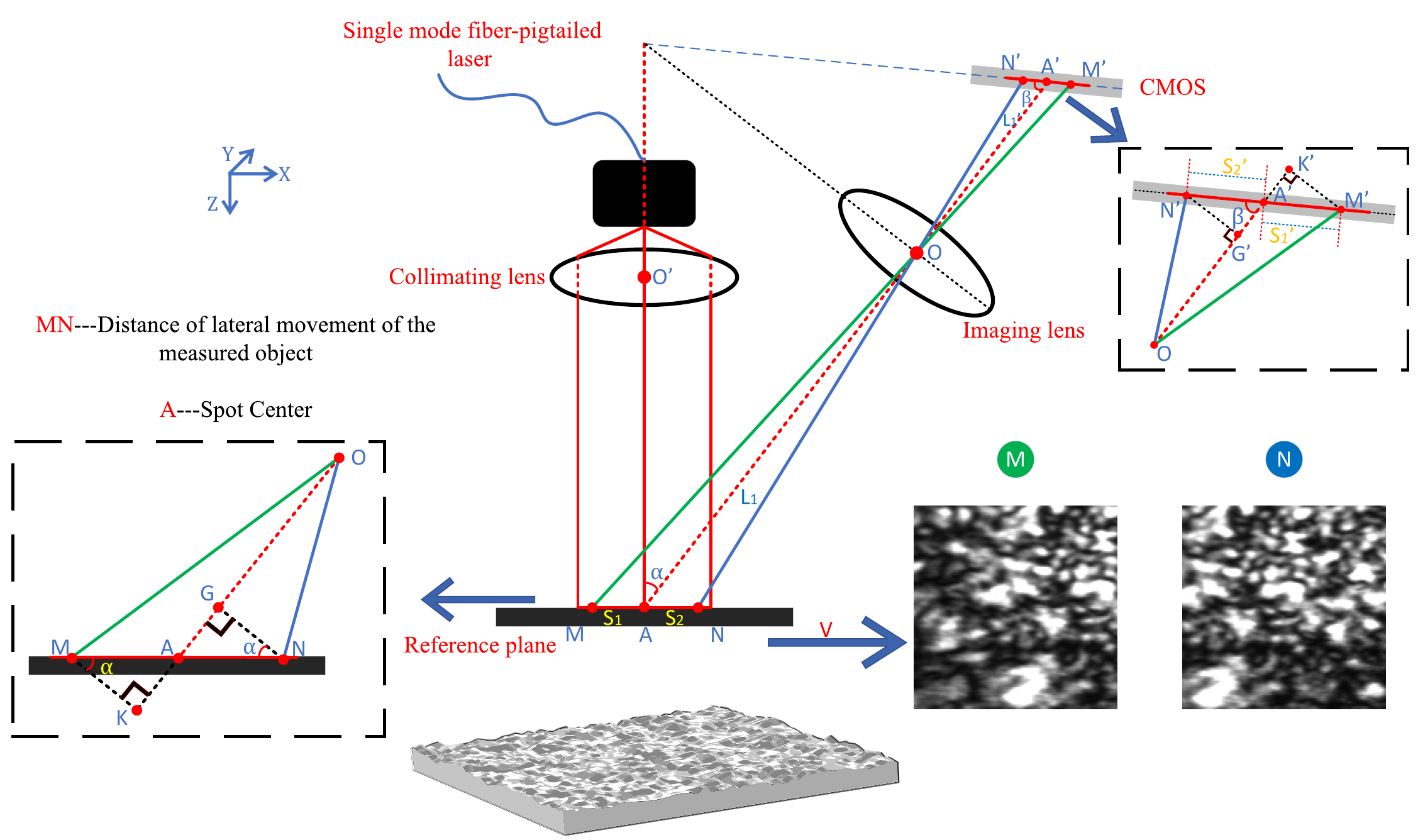}
\caption{Scheme of lateral velocity measurement.}\label{f2}
\end{figure}

Based on the laser triangulation system, the novel theoretical models for lateral velocity measurement are proposed and shown in Fig. \ref{f2} and Eq. (\ref{e2})-(\ref{e4}).

When the measured object moves perpendicular to the direction of the incident beam, the speckle pattern inside the imaging spot shifts correspondingly. Thus, if a position on the object surface covered by the incident beam moves from $M$ to $N$, the speckle pattern captured by the CMOS shifts from $M'$ to $N'$. Based on the beam center, the lateral range of the beam spot is divided into $MA$ and $AN$, and the lateral displacements in both sub ranges are evaluated separately. Firstly, for the movement range $MA$ the right triangle $\triangle OKM$ are similar to $\triangle OK'M'$ and, thus, $\frac{OK}{OK^\prime}=\frac{KM}{K^\prime M^\prime}$ is obtained.

According to the geometric relation of the optical path, $OK=L_1+MA\cdot\sin{\alpha},KM=MA\cdot\cos{\alpha}$, ${OK}^\prime=L_1^\prime+M^\prime N^\prime\cdot\cos{\beta}$, and $K^\prime M^\prime=M^\prime A^\prime\cdot\sin{\beta}$. Among them, $S_1^\prime=M^\prime A^\prime$ is the displacement of the speckle pattern on the CMOS. $S_1=MA$ is the displacement of the object surface covered by the left half of the beam spot.

Based on the imaging law $\frac{1}{L_1}+\frac{1}{L_1^\prime}=\frac{1}{f}$, $\frac{1}{L_1^\prime}=\frac{1}{f}-\frac{1}{L_1}=\frac{L_1-f}{fL_1}$. It can thus be obtained that $L_1^\prime=\frac{L_1\cdot f}{L_1-f}$, where $f$ is the focal length of the imaging lens. With the above relations, the lateral displacement of the surface in the left half of the beam spot is
\begin{equation}\label{e2}
S_1=\frac{S_1^\prime\cdot\left(L_1-f\right)\sin{\beta}}{f\cdot \cos{\alpha}+S_1^\prime\cdot\left(1-\frac{f}{L_1}\right)\cos{\left(\alpha+\beta\right)}}.
\end{equation}

Similarly, for the movement range $AN$, the right triangle $\triangle ONG$ is similar to $\triangle ON^\prime G^\prime$, $OG=L_1-AN\cdot \sin{\alpha}$, $NG=AN\cdot \cos{\alpha}$, ${OG}^\prime=L_1^\prime-A^\prime N^\prime\cdot \cos{\beta}$, $N^\prime G^\prime=A^\prime N^\prime\cdot \sin{\beta}$, $AN=S_2$, and $A^\prime N^\prime=S_2^\prime$. Based on these relations, the lateral displacement of the surface in the right half of the beam spot reads
\begin{equation}\label{e3}
S_2=\frac{S_2^\prime\cdot\left(L_1-f\right)\sin{\beta}}{f\cdot \cos{\alpha}-S_2^\prime\cdot\left(1-\frac{f}{L_1}\right)\cos{\left(\alpha+\beta\right)}}.
\end{equation}

Thus, the total lateral displacement $S_L=S_1+S_2$, and the lateral velocity $v$ can be evaluated by 
\begin{equation}\label{e4}
v=\frac{S_L}{T},
\end{equation}
in which $T$ is the time of the displacement $S_L$.

The measurement range of the lateral velocity is determined by the dimension of the incident beam spot on the measured surface. As a Gaussian beam, the cross-sectional radius of the incident beam spot $\omega\left(z'\right)=\omega_0\cdot\sqrt{1+\left(\frac{\lambda z'}{\pi\omega_0^2}\right)^2}$, where $z'$ is the propagation distance of the beam, $\lambda$ is the beam wavelength, and $\omega_0$ is the waist radius of the beam after passing the collimating lens \cite{svelto1998principles,zhang2023experimental}. With the acquisition period $t$ of the CMOS, the maximum lateral velocity that can be measured is $v_{\rm{max}}=\frac{2\omega}{t}$.

\subsection{Axial distance measurement}\label{subsec3}
\begin{figure}[H]
\centering
\includegraphics[width=1\textwidth]{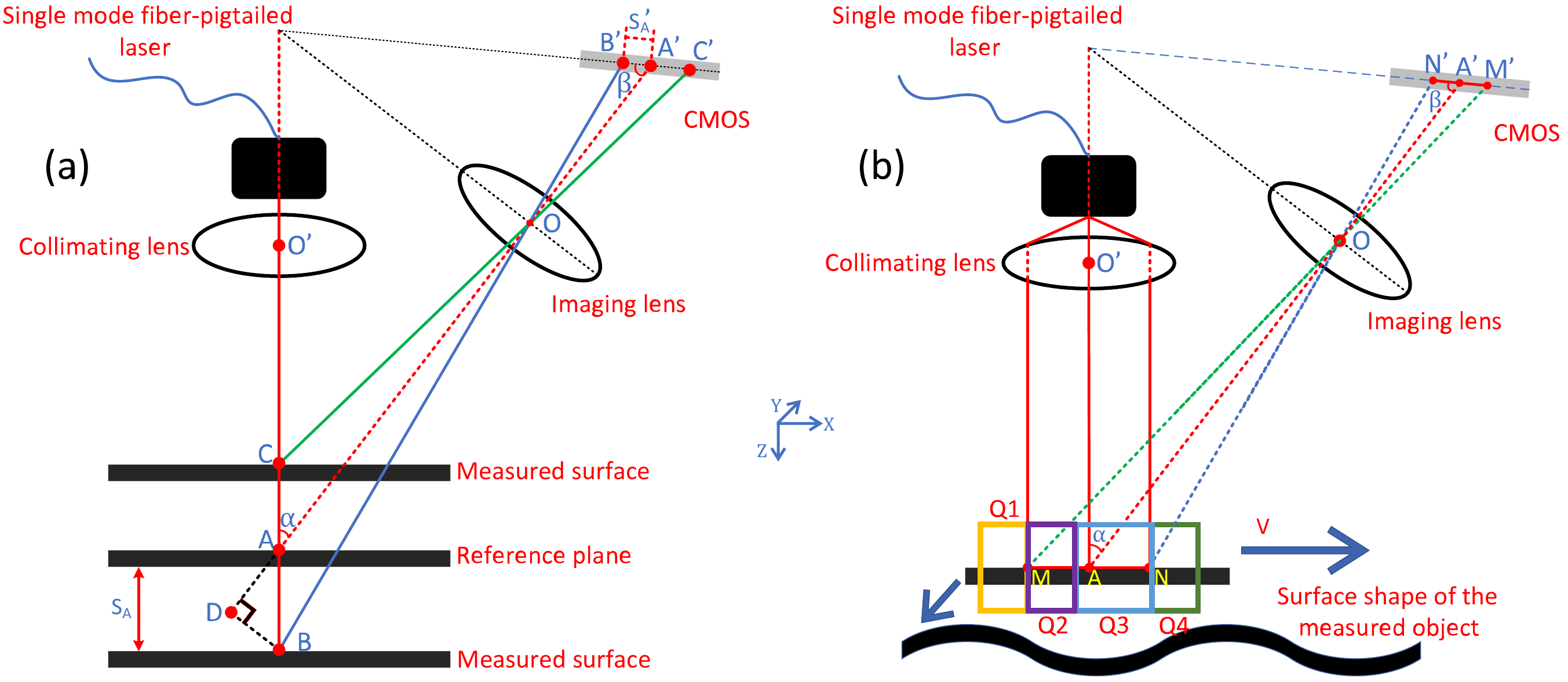}
\caption{Schemes of the axial distance measurement. (a) Measured surface moves along the axial direction. (b) Measured surface moves towards the lateral direction.}\label{f3}
\end{figure}

When there is a position fluctuation of the measured surface along the axial direction, the imaging spot on the CMOS correspondingly moves as well. The axial position (distance) of the measured surface can be estimated by evaluating the position of the imaging spot, and the principle is depicted in Fig. \ref{f3}(a). 

Initially, the laser triangulation system is calibrated by positioning the imaging spot at the center of the CMOS, and the measured surface is on the so-called reference plane. Thus, the axial distance can be evaluated by \cite{zhang2024automatic}
\begin{equation}\label{e5}
S_A=\frac{S_A^\prime\cdot\left(L_1-f\right)\sin{\beta}}{f\cdot\sin{\alpha}\pm S_A^\prime\cdot\left(1-\frac{f}{L_1}\right)\sin{\left(\alpha+\beta\right)}},
\end{equation}

in which $S_A^\prime$ is the position of the imaging spot on the CMOS, and $S_A$ is the axial position of the measured surface. If the surface moves downwards relative to the reference plane, the imaging spot moves to the left on the CMOS, and the denominator takes a $``-"$. Conversely, the imaging spot moves to the right, and the denominator takes a $``+"$.

If the measured surface only moves along the direction of the incident beam, the surface area covered by the beam is constant, and the imaging spot correspondingly moves on the CMOS without changes of speckle pattern. By this case, there is no interference of surface microstructure on the measurement. The axial distance of the measured surface is calculated via positioning the whole imaging spot, which is the common method widely used today. However, as illustrated in Fig. \ref{f3}(b), when the measured surface moves laterally along the direction perpendicular to the incident beam, the speckle patterns of two adjacent light spot images partially change. This not only leads to an uncertainty in the distance measurement but also limits the measurement resolution. 

\begin{figure}[H]
\centering
\includegraphics[width=1\textwidth]{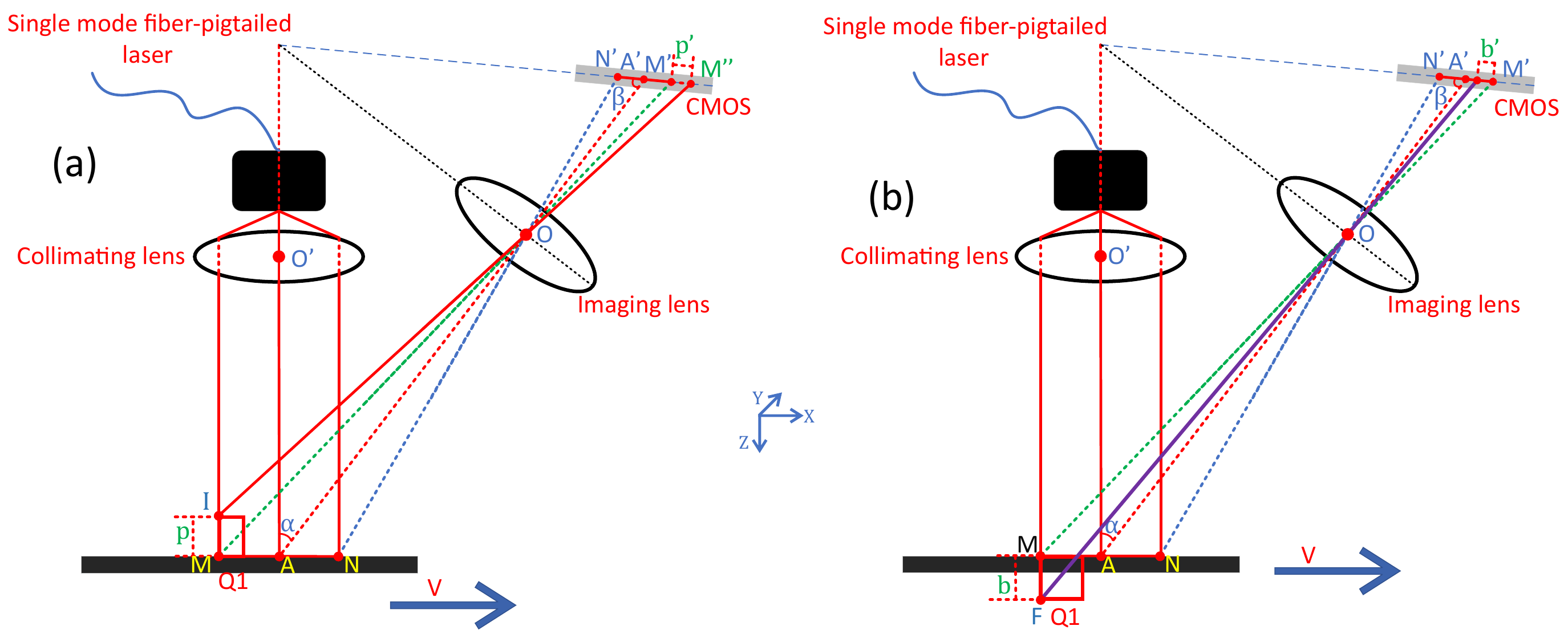}
\caption{Measured surface conditions on the axial distance measurement with lateral motion. (a) The newly entered surface is higher. (b) The newly entered surface is lower.}\label{f4}
\end{figure}

To conquer the above issues, we propose a local edge position based distance measurement method (LEP method). In this method, it can be considered that the surface area covered by the edge of the beam spot varies in real time during the lateral motion. With the acquisition period $t$ of the CMOS, surface $Q1$ with the width of $v\cdot t$ enters the beam spot, and surface $Q4$ with the same width leaves the beam spot. If surface $Q1$ is higher than surface $Q2$, the reflection angle at point $I$ increases causing the right edge of the imaging spot to move towards the right by $p'$, cf. Fig. \ref{f4}(a). The distance can thus be achieved by employing Eq. (\ref{e5}) with the denominator taking $``+"$ and $S_A^\prime=p'$.

When surface $Q1$ is lower than the surface $Q2$, the reflection angle at point $F$ decreases. The right edge of the imaging spot thereby moves towards the left by $b'$, cf. Fig. \ref{f4}(b). Thereby, the distance can be obtained by using Eq. (\ref{e5}) with the denominator taking $``-"$ and $S_A^\prime=b'$. It can be known that on the CMOS the movement of the speckle pattern from $Q2$ and $Q3$ is only caused by the lateral motion of the measured surface, and the edge deformation due to $Q1$ or $Q4$ is related to the distance variation. As a result, the microstructure of the surface can be measured through determining the displacement of the imaging spot edge. This method not only turns the surface microstructure from the uncertainty factor to the measurand, but also increases the lateral resolution from $2\omega$ to $v\cdot t$ by a factor of $\frac{2\omega}{v\cdot t}$.

\subsection{Simultaneous lateral velocity and axial distance measurements}\label{subsec4}

\begin{figure}[H]
\centering
\includegraphics[width=1\textwidth]{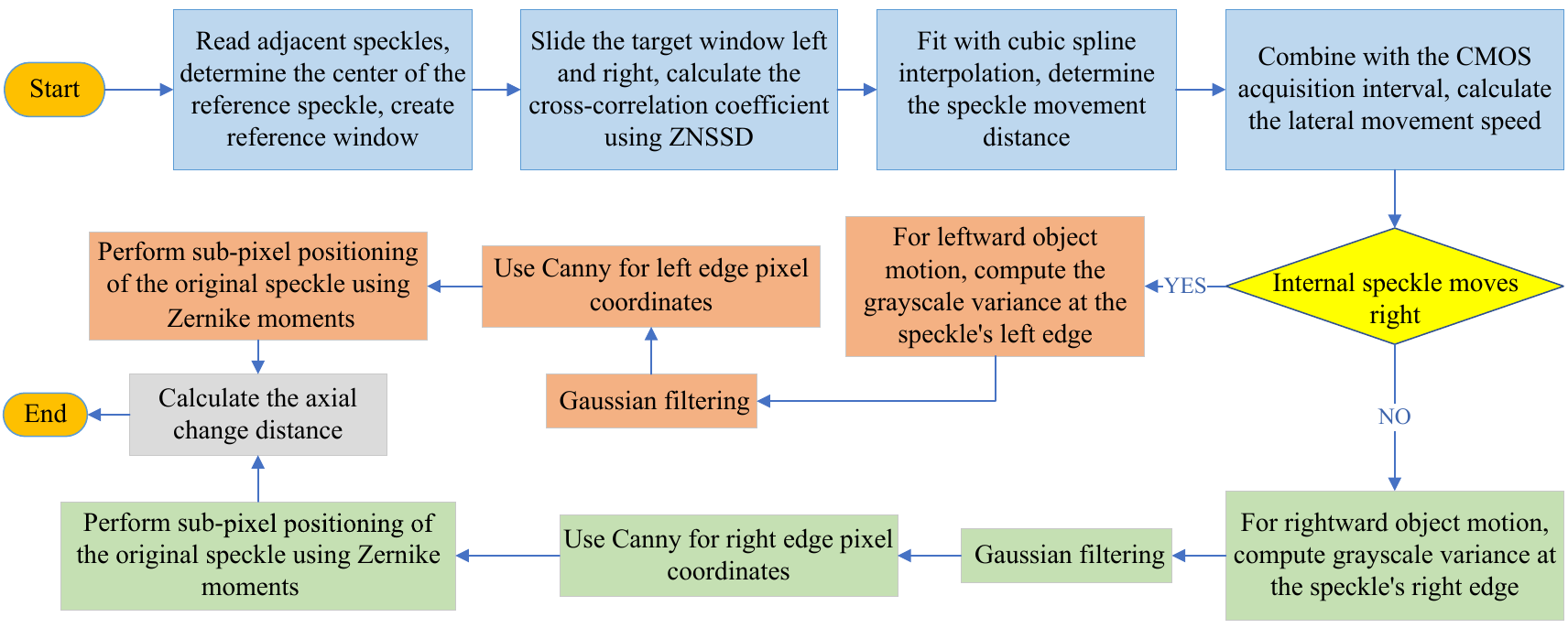}
\caption{Process of the lateral velocity and axial distance measurements.}\label{f5}
\end{figure}

For the lateral velocity measurement, the displacement of the speckle pattern on the CMOS is evaluated by employing digital image correlation. The maximum correlation coefficient is searched for determining the lateral displacement of the speckle pattern and calculated by using the zero mean normalized difference sum of squares function (ZNSSD) \cite{pan2011recent}. Meanwhile, for the axial distance measurement, the Canny operator \cite{canny1986computational} is used to detected the integral-pixel position variation of the imaging spot edge. Based on the integral-pixel position, the Zernike moment method \cite{gao2011zernike} is utilized to evaluate the sub-pixel edge position. The detailed process is illustrated in Fig. \ref{f5}.

In the time resolved speckle image sequence captured by the CMOS, the correlation computation is conducted in each two adjacent images. Firstly, the grayscale centroid method is used to determine the centroid position of the reference imaging spot, and a reference window of $(2r+1)\times(2r+1)$ pixels with radius $r$ centered around the centroid is created. $r=20$ is chosen in this study balancing computation speed, accuracy, and noise influence. A target window with the same size of the reference window is produced for the adjacent target imaging spot. The target window is moved laterally from the left edge to the right edge of the target imaging spot with a step size of one pixel, while the cross-correlation coefficient between the target window and the reference window is calculated by using ZNSSD. The center position difference between the target window with the highest correlation coefficient and the reference window is the integral-pixel displacement of the speckle pattern. Next, the cubic spline interpolation is used to fit the correlation coefficients of integral pixels, and the position of local maximum correlation coefficient around the integral pixel is the sub-pixel displacement of lateral motion. The lateral velocity is then obtained by Eq. (\ref{e4}).

In the axial distance measurement, the lateral displacement of speckle pattern shows the motion direction of the measured surface. As shown in Fig. \ref{f3}(b) and \ref{f4}, if the surface moves to the right, the right edge variation of the imaging spot on the CMOS should be evaluated for the axial distance measurement, conversely, evaluate the left edge. The Gaussian filtering is employed to remove noise and smooth the images. The Canny operator is then utilized to perform integral-pixel localization of the edges in two adjacent imaging spots, and the sub-pixel positions are achieved by using the Zernike moment method. Finally, the average edge displacement is computed and substituted into Eq. (\ref{e5}) to determine axial distance.

\section{Experimental results}\label{c3}

\begin{figure}[H]
\centering
\includegraphics[width=0.9\textwidth]{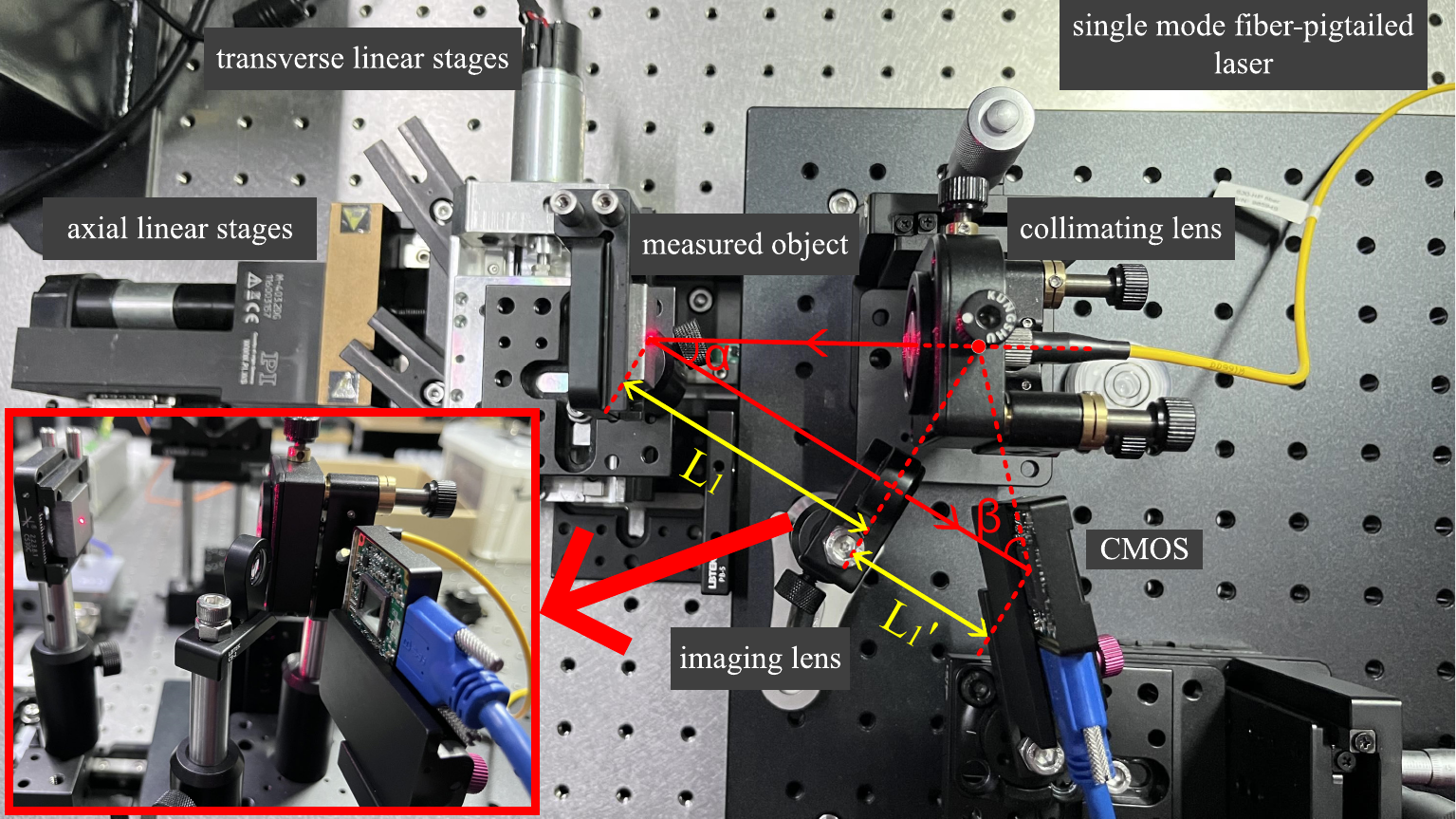}
\caption{Experimental setup of the laser triangulation system.}\label{f6}
\end{figure}

To demonstrate the reliability and accuracy of the proposed laser triangulation based simultaneous two-dimensional velocity and distance measurements, a laser triangulation system is developed, cf. Fig. \ref{f6}.

The incident beam originates from a single mode fiber coupled semiconductor laser (wavelength 658 nm) and vertically illuminates the measured object surface via a collimating lens (aspheric lens, focal length 2 mm). In this case, the radius of the incident beam spot on the surface is about 221 $\upmu$m. The scattered light from the measured surface is imaged by using a CMOS camera (VEN-830-22U3M-M01, DAHENG Corporation) through an imaging lens (aspheric lens, focal length $f=25$ mm). Meanwhile, the receiving angle $\alpha=31.14^{\circ}$, the height of the imaging lens $H=59.27$ mm, and the imaging angle $\beta=46.92^{\circ}$.

\begin{figure}[H]
\centering
\includegraphics[width=0.9\textwidth]{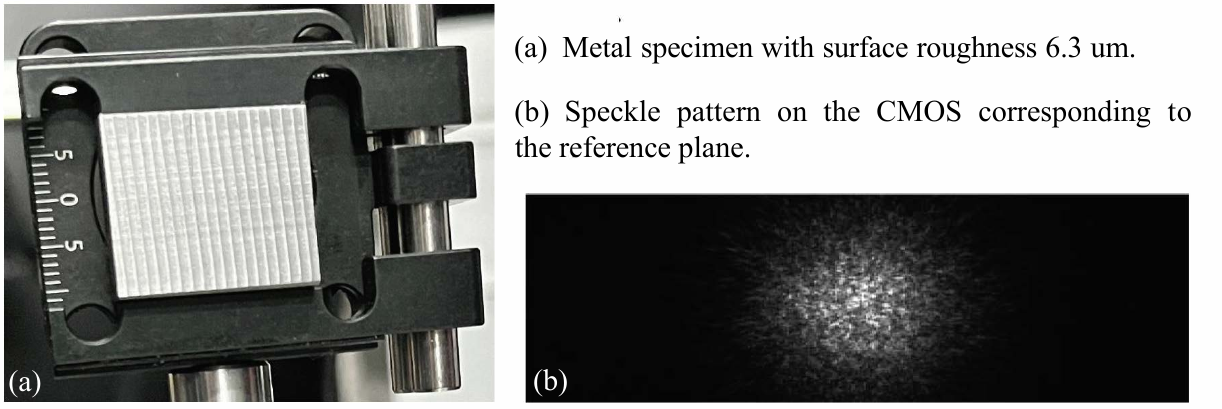}
\caption{Measured object and imaging speckle pattern from the object surface on the reference plane.}\label{f7}
\end{figure}

The measured object shown in Fig. \ref{f7}(a) is a commercial metal specimen with an established surface roughness of 6.3 $\upmu$m and securely fixed on two mutually perpendicular linear motion stage (resolution down to 0.018 $\upmu$m, PI Corporation). The axial stage is used to adjust the measured object surface to the reference plane, and the lateral stage moves the object laterally at different velocities. During the lateral motion of the object, the CMOS captures the imaging spot at a frequency of 100 Hz. The device control and data processing are performed by using MATLAB in a computer
(11th Gen Intel Core i5-11400 2.60 GHz, RAM 32 GB). The reference velocity and distance measurements are conducted by using the lateral motion stage and a spectral confocal sensor (resolution 0.001 $\upmu$m, confocal DT IFC2421, Micro Epsilon Corporation), respectively.

In the experiment, the measured object surface is first moved to the reference plane by using the axial motion stage, which makes the imaging spot at the center of the CMOS, cf. Fig. \ref{f7}(b). The lateral motion stage moves the measured object by the surface length of 22 mm for each measurement. The movement velocity of the lateral stage starts from 1 mm/s and increases by 1 mm/s until it reaches 10 mm/s. 10 measurements are performed at each velocity. Note that the current test velocity is used for principle verification, which can be further increased by employing high-speed imaging devices.

\begin{figure}[H]
\centering
\includegraphics[width=1\textwidth]{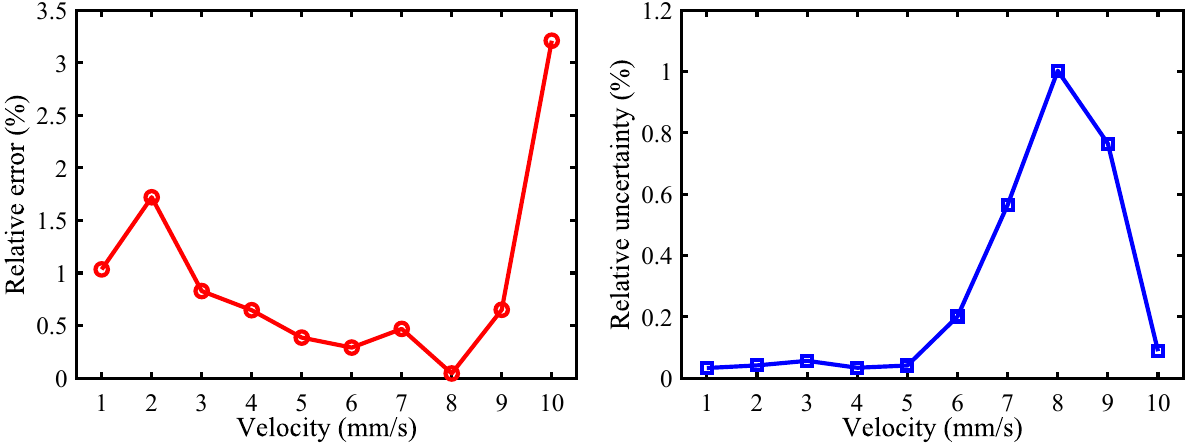}
\caption{Relative error (left) and relative uncertainty (right) of the lateral velocity measurement.}\label{f8}
\end{figure}

The relative error and relative uncertainty of the velocity measurements are evaluated and depicted in Fig. \ref{f8}.  

It can be seen that the evaluations are performed at different motion velocities. The relative errors are in general below $0.83\%$ and can reach $0.04\%$. Moreover, the most relative uncertainties are less than $0.09\%$.

\begin{figure}[H]
\centering
\includegraphics[width=1\textwidth]{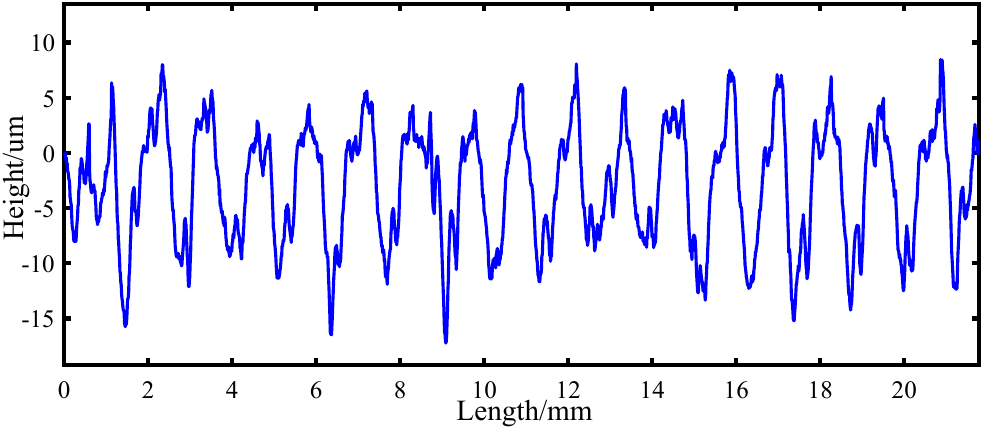}
\caption{Surface shape of the measured object achieved by the axial distance measurement.}\label{f9}
\end{figure}

With the lateral motion the surface shape depended axial distances are measured simultaneously. Compared with the measurement method using the entire spot area, employing the spot edge detection based method the lateral resolution can be increased from 442 $\upmu$m to 10 $\upmu$m by a factor of about 44 in this work. The surface shape is obtained based on the axial distance measurement and illustrated in Fig. \ref{f9}.

\begin{figure}[H]
\centering
\includegraphics[width=1\textwidth]{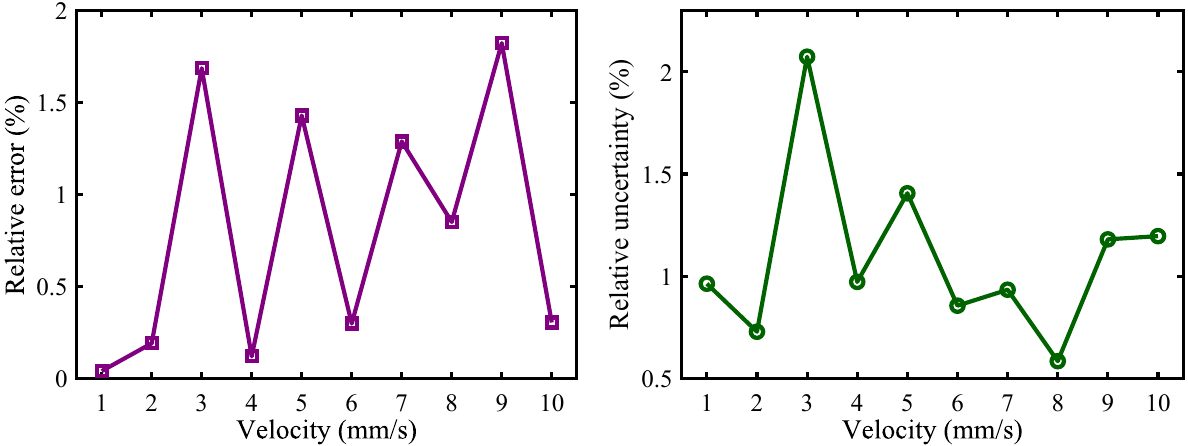}
\caption{Left: relative error of surface roughness. Right: relative uncertainty of surface roughness achieved by the axial distance measurement.}\label{f10}
\end{figure}

The waves of the surface profile are detected successfully and comparable to Fig. \ref{f7}(a). Besides, as the reference measurement the surface is also inspected by using the confocal sensor and the surface roughness of 6.3 $\upmu$m is achieved, which is consistent with the value offered by the manufacturer. The relative error and relative uncertainty of the surface roughness obtained by axial distance measurements of the laser triangulation system at different lateral motion velocities are given in Fig. \ref{f10}.

The relative errors of the surface roughness are shown to be overall below $0.85\%$ and can reach $0.04\%$, where the absolute errors are at the nanometer level. The most relative uncertainties are less than $0.98\%$. During repeated measurements, the motion stage has slight velocity fluctuations resulting in deviations of speckle images, which is a main factor contributing to the measurement uncertainties. Meanwhile, mechanical vibrations and image noise that make speckles vary can also lead to uncertainty components in the measurement results.

\section{Conclusions}\label{c4}

We have introduced and demonstrated the method that makes laser triangulation systems enable simultaneously measuring orthogonal velocity and distance for the first time. The theoretical models for the lateral velocity and axial distance measurements of laser triangulation systems are first established. Furthermore, the measurement algorithms with the ZNSSD, Canny operator, and Zernike moment are developed. Finally, a laser triangulation system is set up, and the experimental measurements on a moving metal specimen are performed compared with the reference measurements achieved by using a commercial precision motion stage and confocal sensor. The experimental results validate that the relative error and relative uncertainty can reach $10^{-4}$. 

The proposed simultaneous two-dimensional velocity and distance measurement method can be applied to all laser triangulation systems. A 3-D measurement can be achieved with a vector decomposition of the lateral velocity measurement. Furthermore, this work could advance the field of dynamic behavior discipline via the applications such as rotating workpiece absolute shape, in-plane and out-of-plane turbo machinery or blade oscillation, gear, spindle play and particle-based fluid velocity measurements, which provides more possibilities for the development of optical measurement technique.

\begin{backmatter}
\bmsection{Funding}
National Natural Science Foundation of China (52205555); Foundation of Liaoning Province Education Administration (JYTMS20230167).

\bmsection{Acknowledgment}
The financial supports of the National Natural Science Foundation of China (project 52205555) and the Foundation of Liaoning Province Education Administration (project JYTMS20230167) are gratefully acknowledged.

\bmsection{Disclosures}
The authors declare no conflicts of interest.


\end{backmatter}


\bibliography{reference}






\end{document}